\documentstyle[11pt,newpasp,twoside,epsf]{article}
\def\edcomment#1{\iffalse\marginpar{\raggedright\sl#1\/}\else\relax\fi}
\reversemarginpar

\begin{document}
\title{Detection of spiral structure of the quiescent accretion disk of IP Pegasi} 

\author{V.V. Neustroev$^1$, N.V. Borisov$^2$, H. Barwig$^3$, A. Bobinger$^3$, 
K.H.~Mantel$^3$, D. \v{S}imi\'{c}$^3$, S.~Wolf$^4$}
\affil{
$^1$ Department of Astronomy and Mechanics, Udmurtia State University,
Izhevsk, 426034, Russia\\
$^2$ Special Astrophysical Observatory, Nizhnij Arkhyz, Russia\\
$^3$ Universit\"{a}ts­-Sternwarte, M\"{u}nchen, Germany\\
$^4$ Max-Planck-Institut f\"{u}r Extraterrestrische Physik, Garching, Germany 		
}

\vspace{1.5 cm}

We present the results of the spectral investigations of the cataclysmic
variable IP Pegasi in quiescence. 
Optical spectra obtained on the 6-m telescope at the Special Astrophysical
Observatory (Russia), and on the 3.5-m telescope at the German-Spanish
Astronomical Center (Calar Alto, Spain), have been analysed by means of Doppler 
tomography and phase modeling technique.

The Balmer Doppler maps show the two bright emitting regions superposed to the 
typical ring-shaped emission of the accretion disk (Fig.~1). 
Brighter of them can be unequivocally contributed
to emission from the bright spot on the outer edge of the accretion disk.
The second spot locate far from the region of interaction between the stream
and the disk particles. This feature was interpreted by Wolf et al.
(1998) as beginning of the formation of spiral arm in the outer disk. 
However the second spiral arm detected by Steeghs, Harlaftis, \& Horne (1997) 
during outburst, on our tomograms is not visible. 
Probably it is hidden in the intensive emission of the bright 
spot. We have modeled emission line profiles and have found a dependence of 
brightness of the spot from an orbital phase (Fig.~2). 
We have detected, that while the spot is on a distant half of the accretion 
disk, it is almost not visible. Apparently, it is connected
to an eclipse of the bright spot by an outer edge of the accretion disk. 
Further we have constructed new Doppler maps, using only spectra obtained during 
phases of minimum brightness of the spot.
Although the used spectra practically do not contain an information about 
the bright spot, in the left side of tomograms can be seen the area of the increased
luminosity (Fig.~1). However its location has varied. It was displaced downwards and 
has taken practically symmetrical position concerning the second region of the 
increased luminosity. We think, that we managed to obtain an image of the 
second spiral arm. Thus, our observations confirm the spiral structure of the
quiescent accretion disk of IP Peg.

\begin{figure}
\centering \leavevmode \epsfxsize=12cm \epsfbox{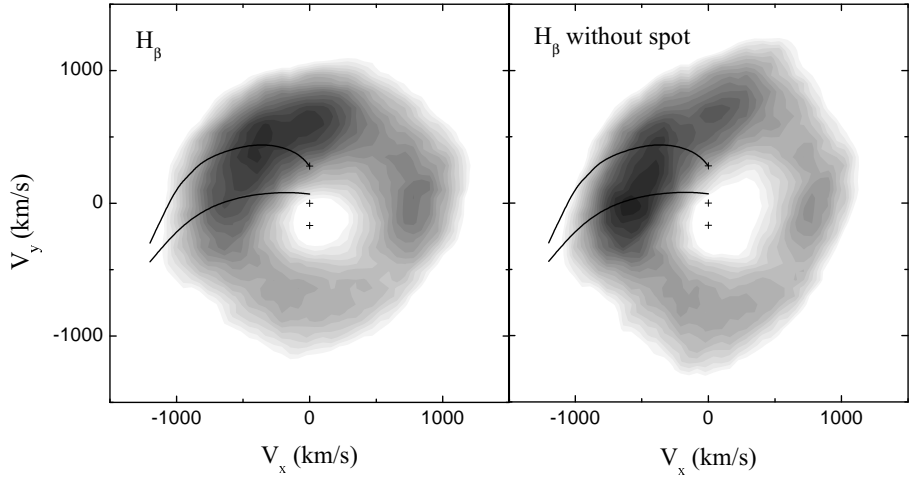}
\caption{Doppler map of H$_{\beta }$ emission. 
Left: Doppler map based on all spectra. 
Right: the map based on those spectra which were obtained 
at minimum spot brightness (for details see text).}
\label{tomogram}
\end{figure}

\begin{figure}
\centering \leavevmode \epsfxsize=11cm \epsfbox{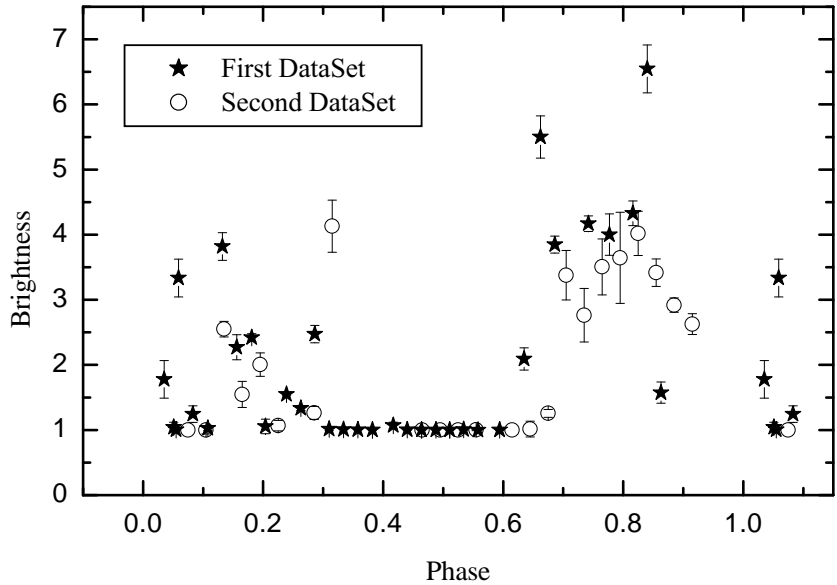}
\caption{The dependence of the spot brightness from an orbital phase}
\label{bright}
\end{figure}

\end{document}